\documentclass[dvips,twoside]{elsart}   
\usepackage{graphicx}			
\usepackage{xspace}                     
\usepackage{ifthen}

\def\KPP{\ensuremath{K_L\rightarrow\pi^+\pi^-}\xspace} 
\def\KME{\ensuremath{K_L\rightarrow\mu^\pm e^\mp}\xspace} 
\def\KMM{\ensuremath{K_L\rightarrow\mu^+\mu^-}\xspace} 
\def\KEE{\ensuremath{K_L\rightarrow e^+e^-}\xspace}
\def\KE3{\ensuremath{K_L\rightarrow\pi e \nu }\xspace} 
\def\KM3{\ensuremath{K_L\rightarrow\pi \mu \nu }\xspace}

\newcommand{\freon}{\ensuremath{\mathrm{CF}_4}\xspace} 
\newcommand{\ethane}{\ensuremath{\mathrm{C_{2}H_{6}}}\xspace}
\newcommand{\isobutane}{\ensuremath{\mathrm{iC_4H_{10}}}\xspace}

\newcommand{\units}[1]{\ensuremath{\,\mathrm{#1}}\xspace}

\newcommand{\cm}{\units{cm}}
\newcommand{\mm}{\units{mm}}
\newcommand{\um}{\units{\mu m}}
\newcommand{\uA}{\units{\mu A}}

\newcommand{\ns}{\units{ns}}

\newcommand{\MeV}{\units{MeV}}

\begin{document}
\newboolean{elstyle}
\setboolean{elstyle}{true}  


\ifthenelse{\boolean{elstyle}}{\begin{frontmatter}}{}
\title{A straw drift chamber spectrometer for studies of rare kaon decays}
\ifthenelse{\boolean{elstyle}}{
  \author[texas]{K. Lang\thanksref{corr}},  
  \author[texas]{D. Ambrose},
  \author[stanford]{C. Arroyo},
  \author[wandm]{M. Bachman},
  \author[irvine]{D. Connor},
  \author[wandm]{M. Eckhause},
  \author[stanford]{K. M. Ecklund},
  \author[texas]{S. Graessle},
  \author[texas]{M. Hamela},
  \author[texas]{S. Hamilton},
  \author[wandm]{A. D. Hancock},
  \author[stanford]{K. Hartman},
  \author[stanford]{M. Hebert},
  \author[wandm]{C. H. Hoff},
  \author[texas]{G. W. Hoffmann},
  \author[stanford]{G. M. Irwin},
  \author[wandm]{J. R. Kane},
  \author[irvine]{N. Kanematsu},
  \author[wandm]{Y. Kuang},
  \author[irvine]{R. Lee},
  \author[texas]{M. Marcin},
  \author[wandm]{R. D. Martin},
  \author[texas]{J. McDonough},
  \author[texas]{A. Milder},
  \author[irvine]{W. R. Molzon},
  \author[stanford]{D. Ouimette},
  \author[stanford]{M. Pommot-Maia},
  \author[texas]{M. Proga},
  \author[texas]{P. J. Riley},
  \author[texas]{J. L. Ritchie},
  \author[richmond]{P. D. Rubin},
  \author[texas]{V. I. Vassilakopoulos},
  \author[texas]{ B. Ware},
  \author[wandm]{R. E. Welsh},
  \author[stanford]{S. G. Wojcicki},
  \author[texas]{S. Worm}
 \\ (BNL E871 Collaboration)
 
  \address[texas]{University of Texas, Austin, Texas, 78712}
  \address[stanford]{Stanford University, Stanford, California, 94305}
  \address[irvine]{University of California, Irvine, California, 92697}
  \address[wandm]{College of William and Mary, Williamsburg, Virginia, 23187}
  \address[richmond]{University of Richmond, Richmond, Virginia, 23173} 
  \thanks[corr]{corresponding author. Email address: lang@mail.hep.utexas.edu}
}{
  \author{
    D. Ambrose$^{1}$,
    C. Arroyo$^{2}$,
    M. Bachman$^{3}$,
    D. Connor$^{3}$,
    M. Eckhause$^{4}$,
    K. Ecklund$^{2}$,
    S. Graessle$^{1}$,\and
    M. Hamela$^{1}$,
    A. D. Hancock$^{4}$,
    K. Hartman$^{2}$,
    M. Hebert$^{2}$,
    S. Hamilton$^{1}$,
    C. H. Hoff$^{4}$,\and
    G. W. Hoffmann$^{1}$,
    G. M. Irwin$^{2}$,
    J. R. Kane$^{4}$,
    N. Kanematsu$^{3}$,
    Y. Kuang$^{4}$,
    K. Lang$^{1,*}$,\and
    R. Lee$^{3}$,
    R. D. Martin$^{4}$, M. Marcin$^{1}$,
    J. McDonough$^{1}$,
    A. Milder$^{1}$,
    W. R. Molzon$^{3}$,\and
    M. Pommot-Maia$^{2}$,
    M. Proga$^{1}$,
    D. Ouimette$^{2}$,
    P. J. Riley$^{1}$,
    J. L. Ritchie$^{1}$,\and
    P. D. Rubin$^{5}$,
    V. I. Vassilakopoulos$^{1}$,
    R. E. Welsh$^{4}$,
    S. G. Wojcicki$^{2}$\and
 (BNL E871 Collaboration)\and
\\
    $^{(1)}$University of Texas, Austin, Texas, 78712,\\
    $^{(2)}$Stanford University, Stanford, California, 94305,\\
    $^{(3)}$University of California, Irvine, California, 92697,\\ 
    $^{(4)}$College of William and Mary, Williamsburg, Virginia, 23187, \\
    $^{(5)}$University of Richmond, Richmond, Virginia, 23173\\ 
    $^{(*)}$corresponding author\\ 
  } 
  \newcommand{\etal}{\emph{et al}}
  \maketitle
}

\begin{abstract}
We describe the design, construction, readout, tests, and performance 
of planar drift chambers, based on 5\mm diameter copperized Mylar and 
Kapton straws, used in an experimental search for rare kaon decays. 
The experiment took place in the high-intensity neutral beam at the 
Alternating Gradient Synchrotron of Brookhaven National Laboratory, 
using a neutral beam stop, two analyzing dipoles, and redundant 
particle identification to remove backgrounds. 
\end{abstract}


\ifthenelse{\boolean{elstyle}}{
  \end{frontmatter}
}{
  \tableofcontents
}

\section{Introduction}
                                                                                
We present a comprehensive description of a two-arm tracking system
consisting of eight planar drift chambers based on 5\mm diameter straws 
designed, constructed, and operated in Alternating Gradient Synchrotron 
experiment 871 at Brookhaven National Laboratory. Following a brief motivation 
for the experiment, we describe mechanical and electronics design considerations, 
important construction details, and three years experience in operating the system. 

\subsection{Physics motivation}

Despite abundant experimental confirmations of the Standard Model of elementary 
particles and interactions, the theory is generally perceived as incomplete, 
or a low energy realization of a more general theory with a full 
symmetry at higher energy. There are indeed many basic questions to which 
answers will have to come from outside the current model. This is the main 
motivation for theoretical and experimental explorations beyond the Standard 
Model. In particular, lepton flavor violation occurs naturally in many 
extensions of the Standard Model; the process \KME is a
sensitive probe of such violation. In general, rare kaon decay 
experiments provided a promising avenue for discovering phenomena outside 
the Standard Model.

The primary goal of experiment E871 was to search for separate lepton number 
violation in the decay \KME. The experiment achieved a 90\% C.L. limit 
of $4.7\times 10^{-12}$~\cite{e871:98ue}, thus probing new interactions in nature 
in the 200-TeV mass scale. To achieve this sensitivity, the experiment ran 
with the very high beam intensities made possible by the Booster at the AGS.

The experiment also made the first observation of the decay \KEE, measuring 
a branching fraction of $8.7^{+5.7}_{-4.1}\times 10^{-12}$ based on four 
observed events~\cite{e871:98ee}. Over 6,000 thousand \KMM events were also 
observed, for a branching fraction of $(7.19\pm 0.17)\times 10^{-9}$, 
reducing the uncertainty for this decay mode by a factor of three
compared to previous attempts~\cite{e871:98uu}.

\subsection{Experimental setup}

Two important experimental elements relevant to this work were the AGS
 high-intensity neutral beam and the E871 two-arm spectrometer. A novel beam 
stop~\cite{worm:95,belz:98} was located in the upstream spectrometer 
magnet to improve downstream acceptance, tracking, and particle identification.

\begin{figure}
\centerline{\includegraphics{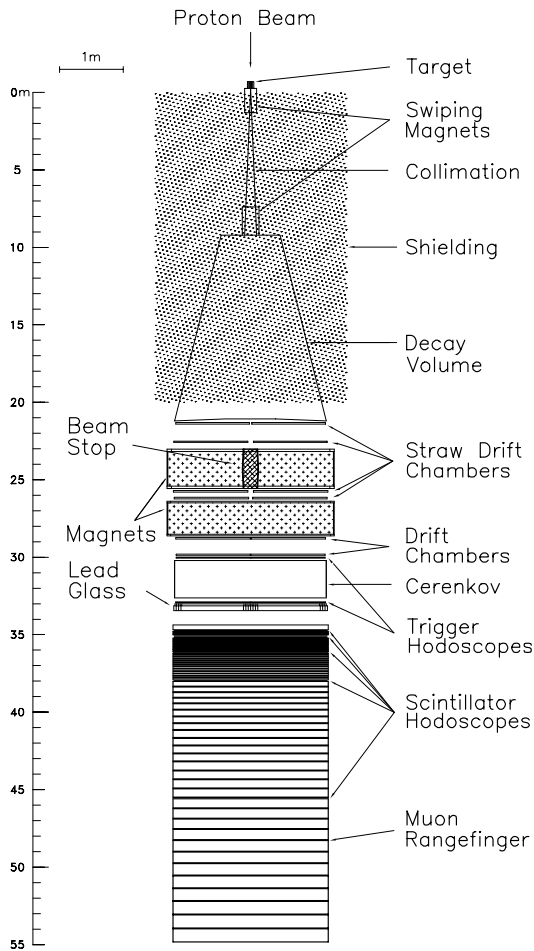}}
\caption{BNL E871 apparatus.\label{E871_ACAD}}
\end{figure}

The essential requirements on the spectrometer were good kinematic
reconstruction and reliable, redundant particle identification,
both to be accomplished in a high rate environment. The spectrometer,
shown in Figure~\ref{E871_ACAD}, was built in the B5 beam line of the
AGS. The neutral beam was produced by an intense primary  24.1\,GeV/c beam
of up to 1.7$\times$10$^{13}$ protons delivered in a 1.6\,s spill
onto a 1.5-interaction-length platinum target.  A system of sweeping magnets
and collimators angled at 3.75$^\circ$ with respect to the proton beam
direction formed a neutral beam with mostly neutrons and kaons at an
approximate solid angle of $4\times 16\,\mathrm{mrad}^2$. Particles emerging
from the 11\,m long evacuated decay volume were tracked and identified in the 
E871 spectrometer. The main features of the apparatus were:
\begin{itemize}
	\item 	two magnets for redundant momentum measurement, tuned to form 
		a ``parallel trigger'' for two-body \KME decays (i.e., charged
		particles traveling downstream of the second analyzing magnets
		triggered the apparatus only if they traveled nearly parallel
		to the neutral beam direction);		
	\item 	a beam-stop placed in the first magnet to absorb 
		the neutral beam
	\item 	redundant finely-segmented fast straw drift chambers, followed 
		by conventional drift chambers in regions of lower rate
	\item 	redundant particle identification of muons and electrons
	\item 	multi-level trigger with fast on-line reconstruction
	\item 	fast custom-designed massively-parallel data acquisition system.
\end{itemize}

The neutral beam was absorbed in a beam stop specially designed and tested for 
this configuration. The effective strengths of the magnetic fields ($+418$ and $-216$\,MeV/c) 
were set such that trajectories of two-body kaon decays emerged nearly parallel 
to the neutral beam direction
downstream of the second magnet~\cite{belz:98}. Such an arrangement simplified 
triggering and provided the first level rejection of the three-body decays. 

The intense primary proton beam produced roughly $2\times 10^8$ $K_L$ decays
per AGS spill, resulting in high hit rates in the upstream straw drift
chambers. Additional rates in chambers resulted from leakage of low energy
particles (mainly charged, but also neutrons and photons) from the
beam-stop. The beam stop shielded the downstream part
of the spectrometer, where rates were substantially reduced, thus minimizing
the probability of pattern-recognition or particle-identification errors.
Rejection of background due to common kaon decays depended crucially on
precise, redundant and low-occupancy tracking.

\subsection{Tracking system requirements}
	%
	%

Since tracking resolution at low momentum is dominated 
by multiple Coulomb scattering, maximal background suppression required that 
the tracking system be low mass, efficient, redundant, fast (i.e., possess 
low occupancy),  and provide good position measurement.  When designing 
experiment 871 the following requirements for the tracking were thus imposed:
\begin{enumerate}
	\item  	minimize multiple Coulomb scattering by limiting the amount of 
                material;
    	\item  	provide tracking redundancy with low rate of wire failures
		and low cell occupancy;
	\item 	provide high hit efficiency for minimum ionizing particles; 
	\item 	sustain good position resolution in high rate environment;
	\item  	assure good vertex reconstruction of two-body decays.
\end{enumerate}
The corollary from the above list are general hardware features desired of 
the system:
\begin{itemize}	
	\item 	high segmentation (i.e., small cell size)
	\item 	fast drift velocity gas
	\item   fast-timing electronics with short pulse tails
	\item	minimal cross-talk
	\item	mechanical robustness.
\end{itemize}
We discuss the design and material choices made to accomplish the above 
requirements in the following sections.

\section{Mechanical design}

\begin{figure*}
    \begin{center}
        \includegraphics[bb= 0 0 352 220]{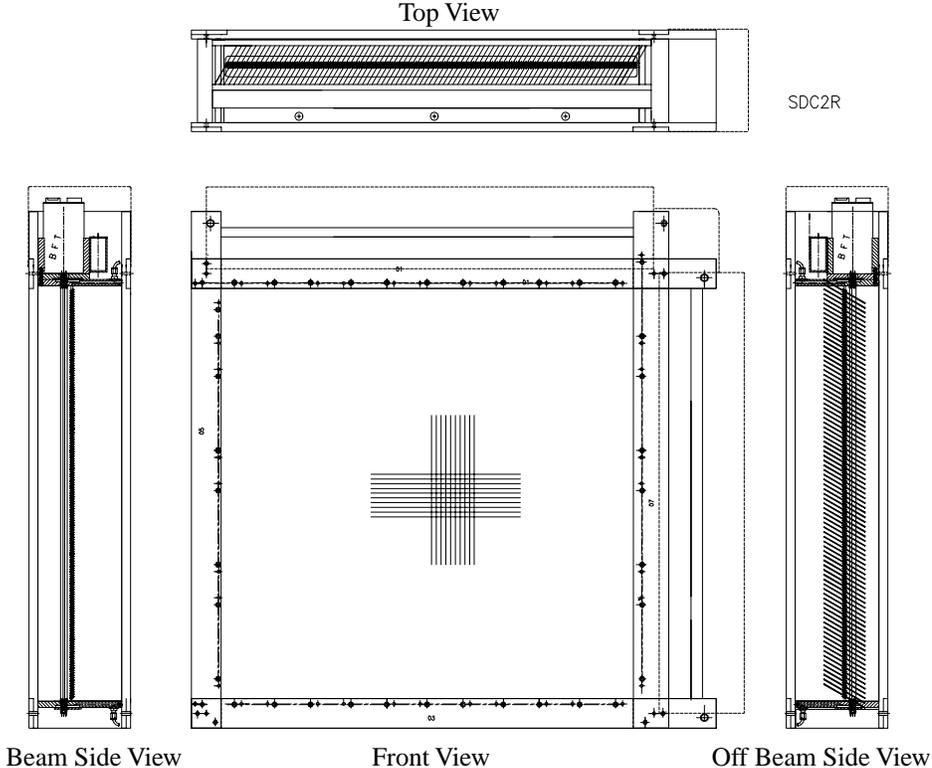}
    \end{center}
    \caption[SDC assembly drawing]{Front, top, and side views of an E871 
	straw drift chamber. The amplifiers are shown in their nominal 
	locations on the top and beam-out sides.}\label{f:sdc-assy}
\end{figure*}

As Figure~\ref{E871_ACAD} illustrates, there were two straw stations between
the decay tank and first spectrometer magnet, and two more between the magnets,
followed by two conventional drift chamber stations downstream of the second
magnet. An assembly drawing of a complete chamber module with $x$ and $y$ views
is shown in Figure~\ref{f:sdc-assy}.

\subsection{Choice of materials}

The technical possibility of making thin-wall tubes suggested the replacement 
of  traditional drift chambers --- with the electric field shaped by wires --- 
by chambers based on a continuous cylindrical cathode surface, or ``straw''. 
The name derives from the manufacturing process, which is similar to that of 
paper drinking straws. Thin, small straws with wall thicknesses of approximately 
30\,$\rm \mu m$ and $5$-mm diameter, were originally used in the vertex detectors of 
several high energy physics experiments at e$^+$e$^-$ 
colliders~\cite{VertexStraws}, as well as in balloon flight 
experiments~\cite{BaloonStraws}. They will be part of the ATLAS detector at 
the CERN Large Hadron Collider~\cite{AtlasStraws}.

Table~\ref{t:sdc-rad} summarizes the amounts of material in an individual 
tracking chamber, expressed in units of radiation length. The traditional 
20\um gold-plated tungsten wire was used for the anode. As shown in the 
table, the total radiation thickness of a typical E871 straw chamber was 
$0.21\% $ of radiation length. 
The thickness of conventional drift chambers 
was very similar.

\begin{table*}
    \caption{Radiation thickness of a straw drift chamber with $3x$ 
	and $2y$ layers (in units of $10^{-4}$ of $X_0$). The straw intra-wall 
	adhesive layer is accounted for by multiplying the number of straw 
	layers (5) by 120\%.}
    \label{t:sdc-rad}
    \begin{center}
        \begin{tabular}{rcccc}\hline\hline
            & mean          &         &            & radiation  \\
            & thickness (cm) & layers & $X_0$ (cm) & thickness (\% $X_0$) \\
            \hline
            Mylar windows   & 0.001270  & 2     & 28.7      & 0.00885   \\
            helium          & 21.434    & 1     & 529888.0  & 0.00374   \\
            Mylar straw     & 0.00798   & $5\times 1.2$\footnote{}
                                                & 28.7      & 0.16683   \\
            copper (1000\,\AA) & 0.000031 & 5   & 1.430     & 0.01098   \\
            \freon (50\%)   &  0.3980   & 2.5   & 9313.0~\cite{tsai:74}
                                                            & 0.01068   \\
            \ethane (50\%)  &  0.3980   & 2.5   & 34035.0   & 0.00292   \\
            tungsten wire   & 0.0000062 & 5     & 0.350     & 0.00886   \\
            epoxy           & 0.000150  & 1     & 35        & 0.00043   \\
            \textsc{Total}  &           &       &           & 0.21329   \\
            \hline
        \end{tabular}
    \end{center}
\end{table*}

The materials most often used for straw walls are polyester (e.g. Mylar~\cite{mylar})
and polyimide (such as Kapton~\cite{kapton}). In our experience Kapton has
the better dimensional stability of the two. Mylar straws installed for 
the 1995 run of E879 showed signs of shrinkage over the course of the running period, 
making the individual straws appear twisted. This motivated our move to Kapton 
for rebuilt 1996 chambers; these chambers displayed no such effect. 
We attribute this difference in stability to the higher water content of 
Mylar relative to Kapton; Mylar became more brittle or ``crinkly'' 
with age --- a sign of dehydration.


Two metals are typically considered for cathode lamination: aluminum 
and copper. Another option is a carbon-doped substrate material
with controlled resistivity. At the time of our construction, doped
Polycarbonate sheets were available and we explored their 
use~\cite{Majewski_polycarbonate}.
However, the thinnest sheets available were 15\um, and since
Polycarbonate is mechanically weaker than Mylar, we did
not pursue this option.
The use of resistive substrate and copper cathodes was motivated by 
effects related to the aging/oxidizing of the aluminum cathode, 
as it had been suggested that a (semi-) conductive base material 
would remedy some of these problems.
Similar comments can be made about Kapton. 
The thickness of Mylar and Kapton can 
be controlled with high precision; thin foils can be as little as 2\um 
thick.  Metalizing Mylar or Kapton up to 1000\,\AA\ is easy and inexpensive. 
Combining Mylar or Kapton with copper provided a strong straw 
base and led to less severe anode aging.

Copper lamination was chosen because of its much better electrical aging 
properties than aluminum. Copper oxide is a semiconductor, which should 
lessen the possibility of cathode field-emission effects, as opposed 
to aluminum dioxide, which is an insulator. Copper cathodes also have 
a relatively high work function, which provides some soft-photon 
absorption at the straw walls.

\subsection{Physical design considerations}

\begin{figure} 
    \centerline{\includegraphics[width=8cm]{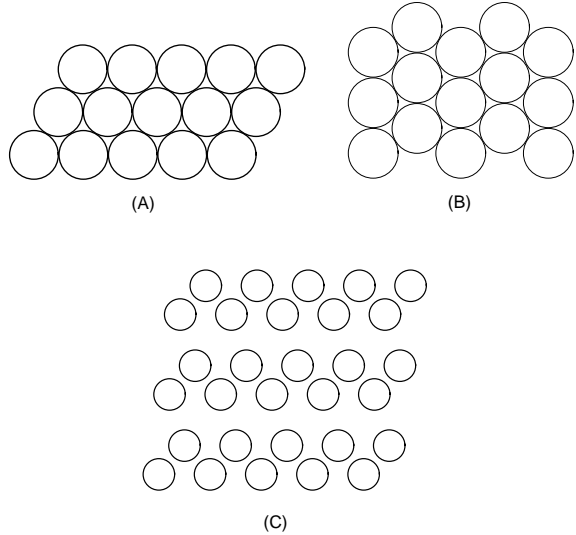}}
    	\caption{(A)~Close-pack, (B)~rotated close-pack, and 
	(C)~staggered geometries. Tracks are vertical in all cases.}
	\label{f:geo}
\end{figure}

We studied several possible planar cell patterns, trying to optimize efficiency
and redundancy while minimizing radiation thickness and construction
complexity. The configurations fall into three categories: close-packed
(Figure~\ref{f:geo}(A)), rotated close-packed (Figure~\ref{f:geo}(B)), and
staggered (Figure~\ref{f:geo}(C)). Close-pack arrangements are bonded together
along the straw, forming self-supporting structures, while staggered designs
require individually tensioned straws.

The self-supporting nature of close-pack designs offers simplicity of design,
since the array needs no external tension. Staggered straws offer greater
flexibility, with a flatter distribution of mean path lengths across the
active area of the chamber, but require tension to keep from sagging.

We studied the relative hit multiplicities of the various geometries via 
Monte Carlo simulation for the acceptance for tracks with angle 
ranges with respect to the neutral beam direction of $-250$ to 250, 20 to 250, 
and 20 to 150\,mr. To maintain high redundancy, we wanted to minimize 
the number of tracks for which two or fewer hits were observed per 
single-view tracking station.
The simulation used the straw's inner diameter of 0.5\,cm as the 
size of the active area of each straw. Simulated E871 tracks were 
projected onto the $xz$ plane and any track crossing a straw's 
active area was considered a hit. 

The studied geometries are shown in Figure~\ref{f:geo} and summarized
in Table~\ref{t:cp}. In the first 
design, a conventional 3-view close-pack, the finite straw-wall thickness 
allowed some 2-hit tracks. The mean straw density was 1.97~straws/cm/view. 
The second design, a ``rotated'' close pack, allowed no 2-hit tracks, 
while some tracks could have as many as 6 hits. The straw density here 
was higher: 2.27~straws/cm/view. Both designs required straws to be 
glued together for mechanical stability. The third design was 
a staggered-straw geometry. This design presented a smoother gas 
path-length distribution and might have allowed replacement of 
individual straws. Moreover the position of the offset straw 
could be placed off-center in order to optimize the chamber for 
a desired angular acceptance. All geometries are shown for three 
measurement views.

\begin{table}[t] \label{t:cp}
	\caption{Comparison of the hit-straw multiplicity, $\bar{n}$,
	and distributions of the number of hit straws, $n$, 
	close-pack(A) or rotated close-pack(B) 
	geometry to staggered geometries of acceptances (C)~-250--250\,mr, 
	(D)~20--250\,mr and (E)~20--150\,mr.  N.N. is the nearest-neighbor 
	distance in cm.}
    \begin{center}
        \begin{tabular}{lccccccc}
        \multicolumn{8}{c}{1.97~straws/cm} \\ \hline\hline
        & & & \multicolumn{5}{c}{Percent $n$-hits} \\ 
        & N.N. & $\bar{n}$ & $n=2$ & $3$ & $4$ & $5$ & $6$ \\ \hline
        A & 0.508 & 2.97 & ~2.7 & 97.3 & ~0.0 & 0.0 & 0.0 \\
        C & 0.683 & 2.97 & 13.1 & 76.5 & 10.3 & 0.0 & 0.0 \\
        D & 0.578 & 2.96 & ~9.7 & 84.3 & ~6.1 & 0.0 & 0.0 \\
        E & 0.651 & 2.97 & 11.7 & 79.4 & ~9.0 & 0.0 & 0.0 \\
        \hline
        \end{tabular}
        \vskip 5mm
        \begin{tabular}{lccccccc}
        \multicolumn{8}{c}{2.27~straws/cm} \\ \hline\hline
        & & & \multicolumn{5}{c}{Percent $n$-hits} \\ 
        & N.N. & $\bar{n}$ & $n=2$ & $3$ & $4$ & $5$ & $6$ \\ \hline
        B & 0.506 & 3.45 & 0.0 & 66.5 & 23.3 & 9.0 & 1.2 \\
        C & 0.635 & 3.47 & 0.1 & 57.0 & 39.1 & 3.8 & 0.0 \\
        D & 0.686 & 3.46 & 0.0 & 57.1 & 40.4 & 1.9 & 0.7 \\
        E & 1.025 & 3.17 & 0.2 & 83.0 & 16.3 & 0.6 & 0.0 \\
        \hline
        \end{tabular}
    \end{center}
\end{table}

Table~\ref{t:cp} compares the close-pack and rotated close-pack 
geometries to staggered straws at the same straw density. All 
configurations with 2.27\,straws/cm/view appear to be hit-equivalent. 
The view densities used here should be considered the two extremes. 
Only a skewed-stagger design, which provides plenty of 
straw-to-straw separation while still limiting the 2-hit 
probability and view density, allows a continuous density adjustment. 

In summary, we found that the rotated close-pack and staggered 
designs with 2.27\,straws/cm/view had similar hit characteristics.  
The mean numbers of hits were 3.45 and 3.46 respectively and neither 
had appreciable 2-hit probability.  Due to the unavoidable 
dead space between adjacent straws, the conventional close-pack 
design allowed about 3\% 2-hit tracks.


The tensioning required of staggered configurations would make chambers 
susceptible to ``creep'', or relaxation of tension, which is expected from 
any plastic stretched beyond its glass transition point. In laboratory testing 
we observed this effect in our Mylar straws. While tensioning or gluing of 
the $x$ (vertical-straw) view would have been unnecessary; $y$ views, 
with horizontal straws, required either tension or structural support 
to maintain shape. Prototype chambers of both close pack and 
individually-tensioned, staggered straws were built and tested in 
cosmic-ray telescopes and in the AGS neutral beam, but the technical 
challenges associated with the individually-tensioned configuration 
led us to choose the conventional close pack design for the E871 spectrometer.

\subsection{Mechanical assembly}

\begin{figure}
  \centering
  \includegraphics[width=8cm]{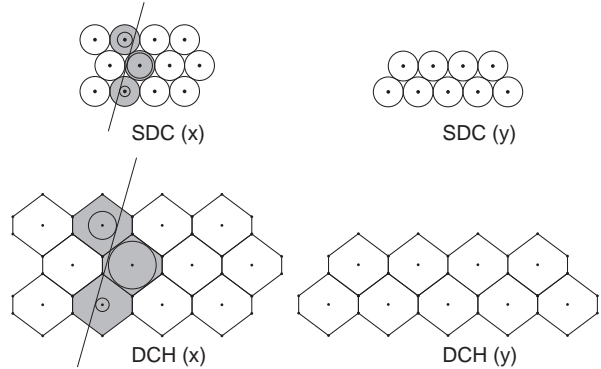}
  \caption[SDC and DC drift cell configurations]{Straw and conventional drift
  cell configurations for $x$ and $y$ views.  Example tracks with (perfect)
  distances of closest approach (DOCAs) are shown.}\label{f:wires}
\end{figure}

Close-pack cell configurations for the E871 straw chambers are shown in 
comparison to those of our conventional drift chambers in Figure~\ref{f:wires}. 
For redundancy in the momentum-measuring horizontal $(xz)$ plane, 
the $x$-measuring views each had a redundant, third plane of drift cells. 
Each drift cell in the straw drift chambers consisted of a 5-mm diameter 
cylindrical copper-plated Mylar or Kapton 
cathode~\cite{straw-manu,straw-manu-2} 
with a 20-micron gold-plated tungsten sense wire~\cite{wire-manu} along 
its axis. The cylindrical cathodes were wound on precision mandrels, 
where they were built up from two 25.4-$\rm \mu m$ layers of material
bonded with a thin layer of 4-8 $\rm \mu m$ of adhesive.
The inner layer had the 0.1-$\mu$m copper layer
vapor-deposited~\cite{vapor-dep} on one side which 
faced the inside of the straw tube.
The original straws in the experiment were made of Mylar. However, 
all but one of the chambers (called SDC4LY) were rebuilt with Kapton for 
the 1996 run period.
Cylindrical brass sleeves provided the mechanical support and 
electrical contact between the straws and the end-plates. Feedthroughs 
pass through the end-plates and into the sleeves. 

\begin{table}
\caption[Straw drift chamber characteristics]{Straw drift chamber characteristics}
\label{t:sdc-stats}
\begin{tabular}{rp{44mm}}\hline\raggedright
diameter	    & 5\,mm			  \\
material	    & Mylar (1995), Kapton (1996)   \\
wall construction   & two 12.5-$\mu$m windings separated by 8-$\mu$m adhesive \\
cathode		    & .1-$\mu$m copper plate, providing 1\,$\Omega/\cm^2$ resistivity \\
anode wire	    & 20-$\mu$m, 4\% gold-plated tungsten \\
mean thickness	    & $4.37\times 10^{-4} X_0$ per layer\\
gas pressure        & 1 atmosphere \\
high voltage	    & 1950--1975\,V \\
max. drift time	    & 25\ns in CF$_4$-C$_2$H$_6$ (1:1) \\
\hline
\end{tabular}
\end{table}

\begin{figure*}
  \centering
  \includegraphics{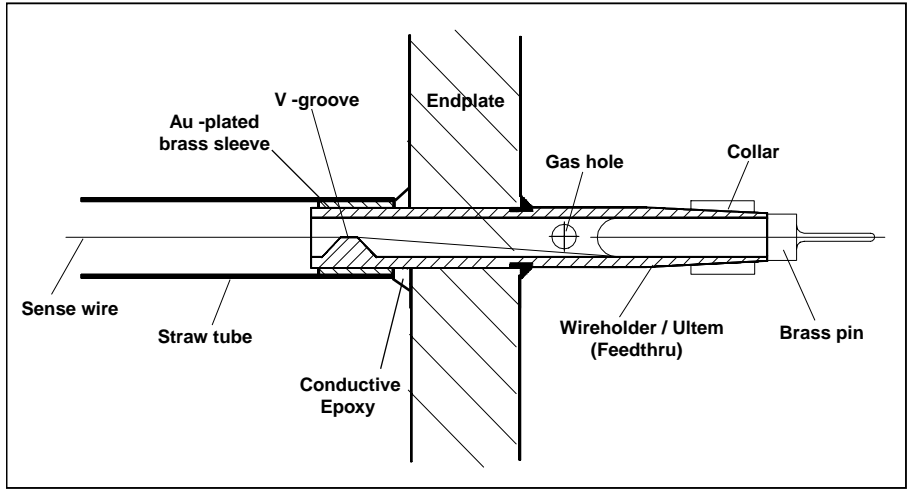}
  \caption[Straw end assembly drawing]{Straw end assembly drawing.}
  \label{f:ultem-assy}
\end{figure*}

\begin{figure}
  \begin{center}
    \includegraphics[angle=90,width=.4\textwidth]{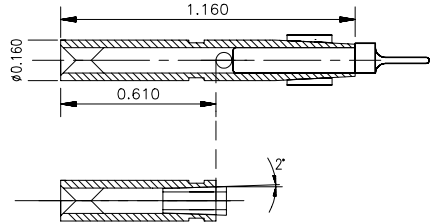}
  \end{center}
  \caption{Standard and truncated ULTEM inserts, with plugs}
  \label{f:ultem-acad}
\end{figure}

\begin{figure}
  \begin{center}
    \includegraphics[angle=90,width=.4\textwidth]{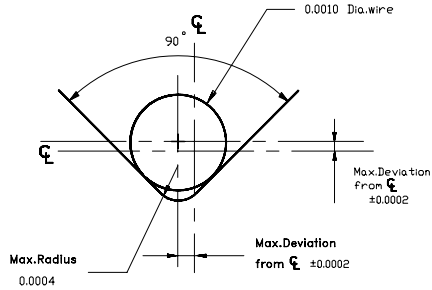}
  \end{center}
  \caption{Wire-positioning V-groove}\label{f:ultem-groove}
\end{figure}

\begin{figure*}
  \centering
  \includegraphics[ angle=90, width=\textwidth]{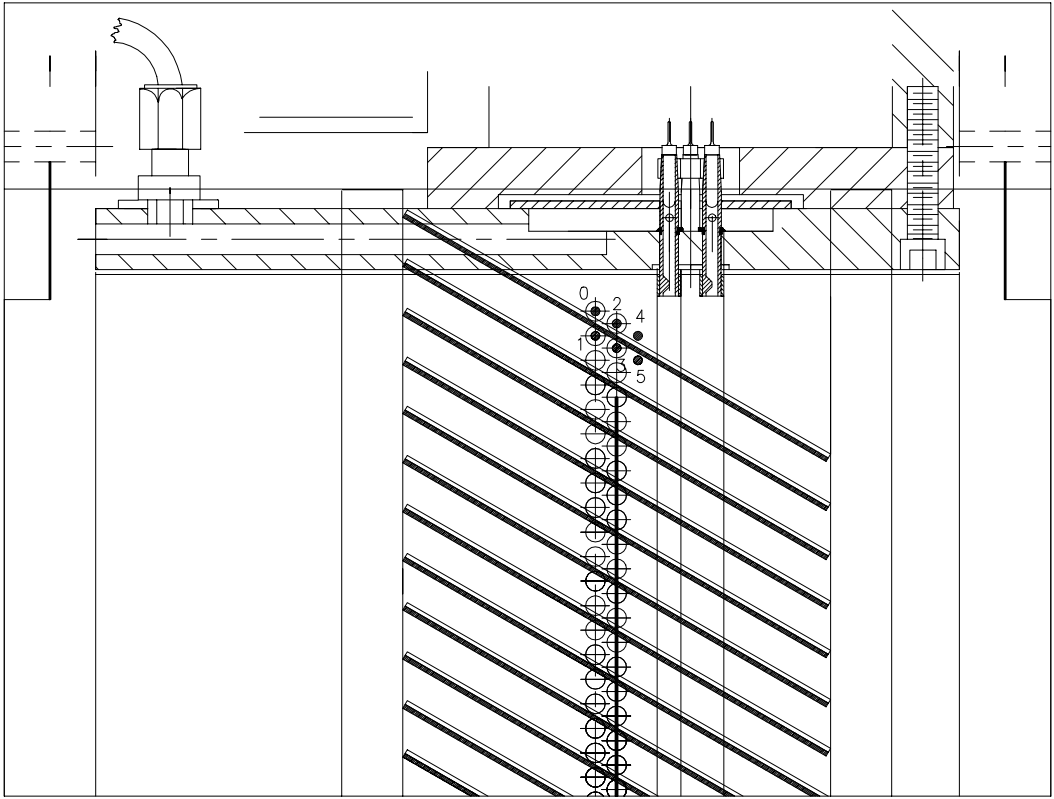}
  \caption[SDC chamber assembly]{SDC chamber assembly.}\label{f:sdc-corner}
\end{figure*}

Individual straws were bonded into closely-packed arrays whose cross
section is shown in Figure~\ref{f:wires}. Sense wires were precisely
positioned within the straws by molded Ultem \cite{ultem-manu} inserts, which
also acted as gas feedthroughs.  Ground contact between the copper cathodes and
the aluminum chamber frames was made through gold-plated brass sleeves, which
were press-fit onto the Ultem feedthroughs and silver-epoxied into the straws. 
More silver epoxy formed the electrical contact between the sleeves and the
chamber frames, while also providing a gas seal. 
Figures~\ref{f:ultem-assy},\ref{f:ultem-acad},\ref{f:ultem-groove} show 
a cross-section assembly drawing of the inserts, sleeves, and end-plate gas
manifold. A more detailed end-plate assembly for one of the chambers, including
amplifier-card mounting and gas connections, is shown in
Figure~\ref{f:sdc-corner}. The straw drift chamber characteristics are
summarized in Table~\ref{t:sdc-stats}.

The feedthrough assembly served a number of functions, including positioning, 
tensioning, and electrically isolating the anode wire from the end-plate. 
It also supported the electrical contact of the anode wire to the outside 
signal and high voltage, and provided gas flow into or out of the straw. 
A sketch of the end-plate and feedthrough assembly is shown 
in Figure~\ref{f:ultem-assy}. 

The wire was positioned at the center of the straw at the feedthrough with a
precision of about 25\,$\rm \mu m$ using a V-shaped groove molded into the feedthrough
near one end (Figure~\ref{f:ultem-groove}). Tension was maintained with a brass
pin inserted and clamped into the slotted Ultem end opposite the V-groove. The
pin diameter, about 50\,$\rm \mu m$ smaller than the inside diameter of the
feedthrough, allowed the anode wire to extend along the pin. The wire was
compressed between the brass pin and the (slotted) feedthrough wall with an
Ultem ring, which slid over the outside of the feedthrough.The pin end of the
feedthrough, the pin, and the compression ring extended outside the chamber
end-plate.

The Ultem material had a resistance of at least 100\,M$\Omega$ 
through its wall thickness of 0.10\,inches. Gas was supplied to 
the straw tubes in parallel through a gas manifold milled into 
the end-plates. From the manifold, the gas entered the interior 
of the Ultem feedthrough via a pair of holes in the feedthrough 
walls. 



Since the straw assembly was essentially irreversible, 
we did a test assembly of the complete chamber frame, 
including end-plates and corner blocks. This was then dismantled 
for installation of straws. All straw-end pieces were deburred and cleaned 
in an ultrasonic bath. Prior to assembly,
holes with critical clearance, such as the 
Ultem clearance holes in the end-plates, were tested and cleaned 
by inserting the requisite gauge pin. Most gas fittings were 
installed in the frame pieces prior to assembly, since 
the assembled frame had insufficient clearance for installation. 
With the exception of the first chamber built (called SDC2R), which 
had a stainless steel frame, all frame pieces were aluminum.


The straws, of Mylar or Kapton, were shipped with Tyvek~\cite{Tyvek} 
liners, in either PVC tubes or wooden crates. Thus the first step in 
the straw preparation was unpacking, de-lining, and inspection. 
Straws with visible inner surface irregularities (dents or lumps 
of glue) were rejected. Following 
a previous experiment~\cite{McDonald_straw_cutting}, 
we constructed a simple straw cutting method consisting of 
a precision bored Al block, with a hole diameter about 
25\,$\rm \mu m$ larger than the outer diameter of straws,
with a transverse slit cut through the bore. Cutting a straw 
only required insertion into the bore and sliding a razor into 
the slit. 
 
        
\begin{table*}
  \caption{Adhesives and encapsulants used in the E871 straw chambers}
  \label{t:glues}
  \begin{center}
    \begin{tabular}{llcc}\hline\hline
    application         & name              	& pot life  & cure time\\ \hline
    straw-insert bond   & EPO-TEK EP110 silver epoxy & & \\
    insert-frame bond   & EPO-TEK 410E silver epoxy  & & \\  
    straw-straw bond    & EPO-TEK 301-2 epoxy   & 24\,h     & 48\,h     \\
    ULTEM/frame seal    & TRA-BOND 2143D epoxy  & 75\,m     & 18\,h     \\
    outer ULTEM seal    & DOW 3110 two-part RTV & 165\,m    & 6.5\,h    \\
    \hline
    \end{tabular}
  \end{center}
\end{table*}
	
Straws were installed in each $x$ and $y$ view of each module separately. 
Before the two views were combined, their straws were laminated, potted, 
and structurally glued to the view sub-frame. Straws were held together 
within an array through spot application of Epo-Tek 301-2, a very liquid 
epoxy with an 8-hour pot life and two day cure~\cite{epo-tek}. After 
securing the straws precisely in place with specially-made brackets mounted on 
the chamber frames, epoxy was applied in small ($\sim 0.5$\,mm-diameter) 
drops to 4--6 points along each tangent between straws. A gas-pressure 
dispenser was used, with tubular steel tips which allowed inner straws 
to be reached. 

At this point the two views fit loosely in their frames. Bolting and 
pinning the $x$ and $y$ views together completed and squared the 
module frame, making it possible to bond the straw arrays to the 
frames by applying Tra-Bond 2143D epoxy~\cite{tra-con} around each 
insert outside the chamber end-plates, but still inside the gas 
manifold. Electrical contact between the straw inner surface 
(via the gold-plated brass `sleeve') was made by filling the gap 
between straw/sleeve end and inner end-plate surface with EPO-TEK 
410E silver-filled epoxy~\cite{epo-tek}. Although before 
installation individual straw assemblies were tested for leaks, 
the silver epoxy should form an additional gas seal; the 2143D 
epoxy forms the manifold-straw seal. The frame assembly was completed
by gluing down of the manifold cover with a bead of silver epoxy 
joining the aluminum cover 
to the end-plate and filling the 2143D epoxy around each ultem insert. 
 

Wires were installed after chamber assembly. We used 20-$\mu$m, 
gold-plated tungsten wire. The wire was threaded through a blunted 
sewing needle, fed through the top insert, and lowered down 
the straw and out the bottom insert. To provide working 
clearance and hold the wire against the interior v-notch of 
the insert, the wire was deflected by a horizontal bar suspended 
below the chamber. A gold-plated brass pin was inserted into 
the top insert and clamped in place with an Ultem collar. Then 
tension was applied to the wire by clamping a 70-g mass below 
the bottom insert and deflecting bar (some tension was lost to 
friction at the bar and at the Ultem notch). Wire tension was 
locked in by installing another brass pin in the bottom insert, 
again with a clamping Ultem collar. Because of the awkward 
position and close-pack clearances of this step, we designed 
a special tool for pin installation. Once we were satisfied 
with the wire installation, a drop of epoxy (301-2) at each 
end ensured the pins and wire ends stayed in place.
 

We checked wire tensions by measuring the resonant frequency 
of a driven wire in a constant magnetic field. 
A custom device~\cite{WireTensionMonitor} 
performed the frequency measurement and provided the driving 
frequency, while a permanent magnet held near the wire center 
supplied the field. Any wires not satisfying minimum tension 
or standoff limits were removed and replaced.


\section{Readout electronics}

\begin{figure*}
  \begin{center}
    \includegraphics[width=\textwidth]{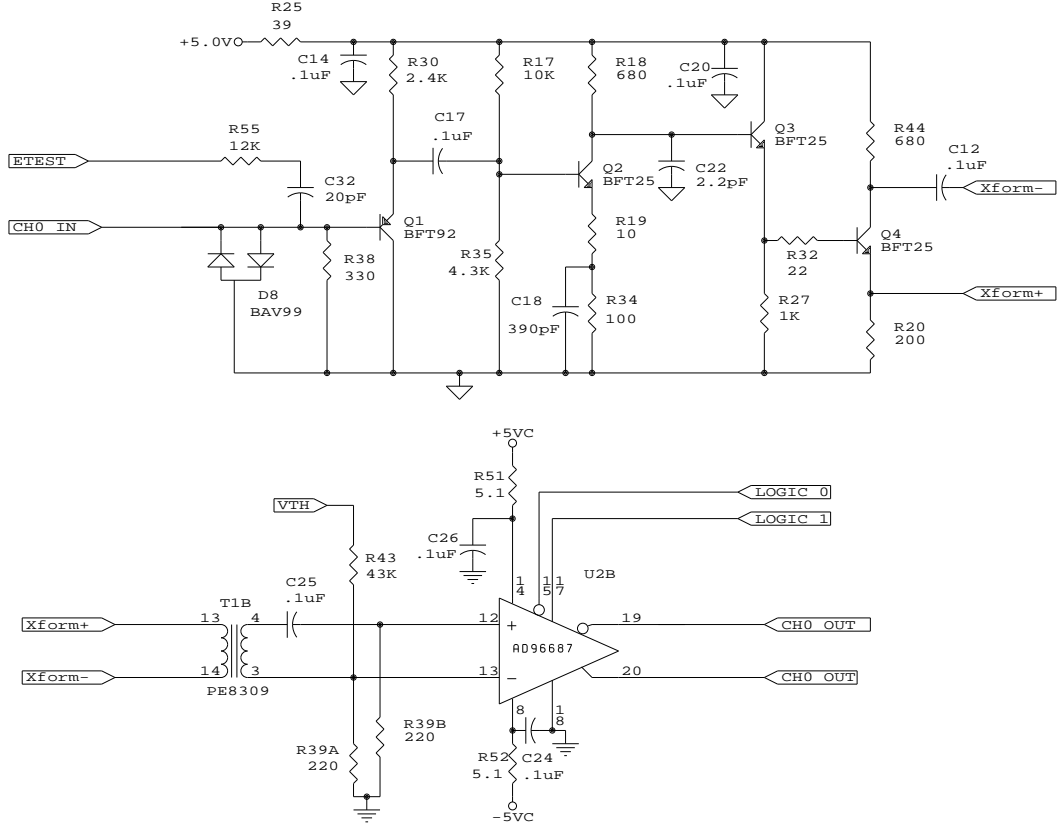}
  \end{center}
  \caption{Amplifier schematic for the straw drift chambers.  
  A single amplifier and discriminator circuit is shown.}
  \label{f:sdc-amp}
\end{figure*}

Signals from the straw sense wires were amplified and discriminated 
on 6-channel boards directly mounted on the straw end pins. The 
amplifier circuit (Figure~\ref{f:sdc-amp}) is a 
four-transistor design, decoupled from the high-power discrimination 
stage by a 1-to-2 transformer \cite{graessle:95}. The gain was about 
20\units{mV/\mu A} with less than 1.5\% cross talk, and a typical 
operating threshold of 1.5\units{\mu A}. The digitized signal was 
converted to a 30-ns pulse and sent over a 333-ft long, 500-ns delay 
line by a 96-channel driver board, to be recorded by a 6-bit, 
1.75-ns least-count TDC.

Since particle positions were determined from the drift 
times of ionization electrons, position resolution was directly 
affected by the precision of the drift-time measurement. The high 
particle flux expected in experiment 871 required a small cell 
size to minimize occupancy, with a high drift velocity gas for reduced 
dead time. 

\begin{figure}
    \begin{center}
        \includegraphics[width=8cm]{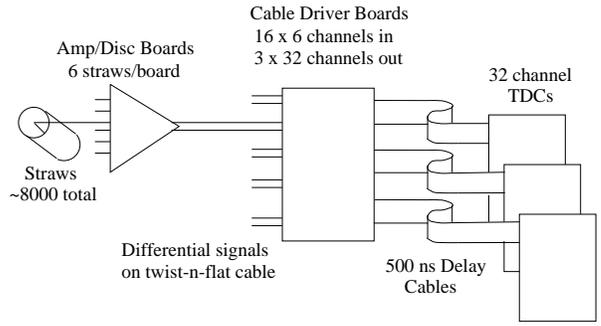}
    \end{center}
    \caption{E871 straw chamber readout}
    \label{f:ele-block}
\end{figure}

The small cell size of the straw chambers called for tight 
packing of front-end channels, increasing the possibility of 
cross talk. Our plan to use a \freon drift gas, meanwhile, 
implied small pulse heights, requiring a high gain front-end 
amplifier. In addition, the front-end electronics were required 
to dovetail with existing equipment: 32-channel gray code TDCs 
with 1.875\ns least count~\cite{r:fast-tdc} coupled to the 
front-end electronics through 97-conductor (32 ECL channel) 
500\ns delay lines. Front-end components designed and built 
specifically for E871 included 6-channel amplifier/discriminator 
boards, which mounted directly onto the chamber frames, and 
96-channel cable driver boards. A block diagram of the straw 
chamber electronics setup is shown in Figure~\ref{f:ele-block}.

Initial designs of the front end amplifier based on an 
integrated-circuit amplifier proved costly (more than 12 dollars 
per channel), while an ASIC-based design missed our time window. 
We therefore pursued a much less expensive option (less than five 
dollars per channel), based on a four-transistor discrete amplifier 
design. Our initial design, with high-voltage distribution and 
the analog amplifier circuit on the chamber-mounted board and 
the discriminator on a separate driver board, taught us that 
twist-n-flat cable made an excellent antenna. The discriminator 
stage was subsequently moved onto the chamber-mounted board.

The final 6-channel amplifier/discriminator layout consisted of 
high-voltage, analog, and digital sections coexisting on one 
four-layer board. Copious grounding and surface-mounted components 
eased the tight space constraints, while SIP sockets facilitated 
mounting the cards directly onto the straw end pins. The bias 
high voltage and front-end amplifier low-voltage power, along 
with odd- and even-channel pulser signals for rudimentary 
electronics diagnosis, were delivered by chamber-mounted busses. 
The digital half of the amplifier boards were supplied by 
a separate cable bus.

\subsection{Early testing}

Early bench tests with simulated straw pulses reported gains of 
$(22.\pm 1.2)\units{mV/\mu A}$. Stable operation was achieved 
at thresholds as low as 0.9\,\uA. We found that DIP-package 
multi-channel transformers in our original amplifier design 
caused too much cross talk. After replacing these with 
single-channel packages, cross talk between adjacent channels 
was reduced to 1.5\%.

The front-end amplifiers were also tested on prototype straw 
chambers. During cosmic ray tests the amplifiers again 
operated at thresholds below 1\uA, with overall straw efficiencies 
measured above 97\%. Single-wire position resolution for the 
cosmic ray tests was 120\um for the Ar-ethane (50-50) gas-fill.

\subsection{Full-system performance}

During actual data taking for E871, the complete 6500-channel 
straw chamber readout system performed nearly as well as in 
tests. Even in the noisy environment of a high energy physics 
experiment, all chambers operated at or below a 1.3-$\mu$A 
threshold with a mean efficiency of 96\%.

We experienced a significant board failure rate due to cold 
solder joints on the amplifier cards; a test stand was built 
to diagnose and repair these faults.

The entire mechanical design  and the front-end electronics system 
has subsequently been copied for use in the Fermilab FOCUS experiment 
(FNAL 831)~\cite{fnal831}.

\section{Gas system}

The gas delivery system was built to handle multiple gas 
mixtures, applying both high- and low-pressure filtering. 
Its purpose was to provide an uninterrupted flow of clean 
gas mixture, with a constant ratio of its components while 
maintaining stable pressure inside and outside the straw 
volume. Additional attention was paid to purifying the gases, 
since impurities in the gas mixture can catalyze straw etching 
or affect the attachment coefficient, charge-carrier diffusion, 
or drift velocity. 

In order to minimize multiple scattering effects, we flowed 
helium through the chamber volumes surrounding the straw tubes, 
while helium-filled bags occupied the spaces between chambers. 
The gas system had to regulate the helium flow so that 
differential pressure across walls was restrained and flow 
was sufficient to prevent accumulation of flammable gas in 
the helium-filled volumes.

During normal operation the detectors and a large part of the gas system were
in a high radiation environment. It was critical to have remote access to the
status of the gas system. Supply pressures, mixture flow rates and ratios,
input and output pressures, and differential pressure across the straw walls
were continuously monitored and periodically inserted into the data stream.


\subsection{Choice of gas}

The high beam intensity in E871 necessitated not only increased 
segmentation of the upstream chambers, but also the use of a gas 
with a higher drift velocity, thus reducing the overall occupancy.

\begin{figure}
\centerline{ \includegraphics[bb= 0 20 283 240]{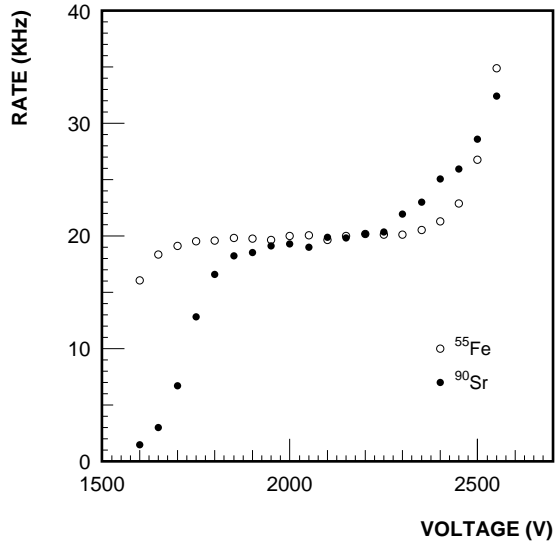} }
\caption{Plateau curves for 
\freon-\ethane (50:50) in response to two different sources.} 
\label{f:freon-plateau}
\end{figure}

\begin{figure}
  \centerline{\includegraphics[bb= 0 20 283 240]{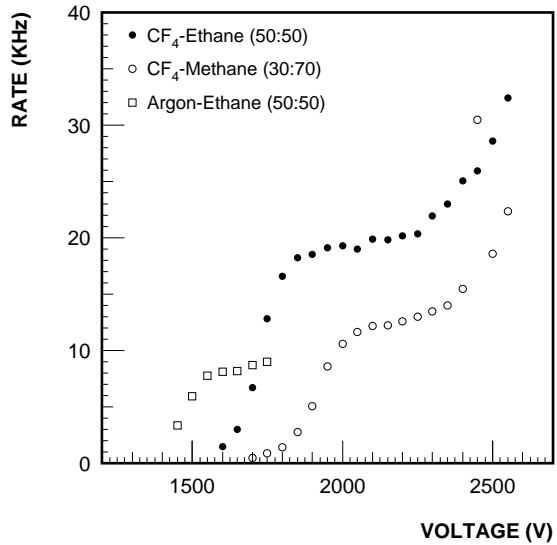}}
  \caption{Plateau curves for various gases with a $^{90}$Sr source. 
  (The source was attenuated for the methane and Ar-Et data.)} 
  \label{f:plateau-comp}
\end{figure}

Most fast drift mixtures considered by various straw/chamber groups 
have been based on tetrafluoromethane~\cite{Vavra:1993cy}, 
\freon (also known as Halocarbon 14). Our decision to use 
\freon-\ethane (50:50) was based primarily on studies performed 
by Heinson and Rowe~\cite{heinson:92}. They list several advantages 
of an ethane mixture over the commonly-used \freon-\isobutane (80:20), 
including a 10\% higher drift velocity, safer handling, smaller radiation 
length, insensitivity to soft photons --- they report the ethane mixture 
to be 25\% less sensitive than isobutane. Perhaps most notably, they 
also found the \freon-\ethane mixture to be less expensive and to 
permit a 200-V lower operating voltage. Plateau curves observed in 
the laboratory with a prototype chamber for this gas are shown in 
Figure~\ref{f:freon-plateau} for two different radioactive sources. 
We chose a nominal operating voltage of 1975\,V. Figure~\ref{f:plateau-comp} 
compares the response of this gas to some other potential mixtures.

\begin{figure*}
\centerline{ \includegraphics{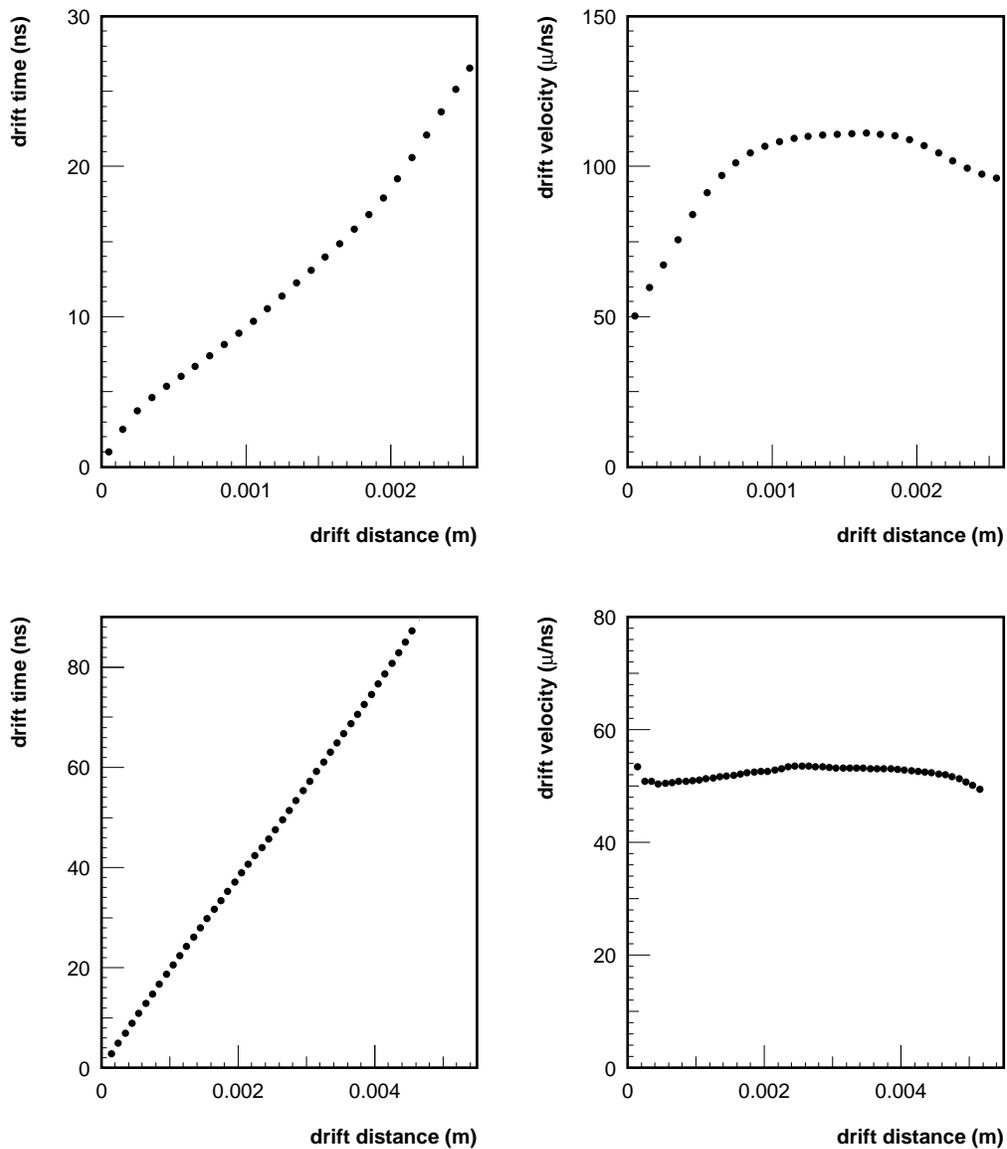} }
\caption{Drift time and velocity for straws with \freon-\ethane (50:50) 
(upper plots) and conventional chambers with Ar-\ethane (50:50) (lower plots) }
\label{f:drift-t}
\end{figure*}

Figure~\ref{f:drift-t} shows the drift time for this mixture as 
a function of  distance of closest approach, along with the drift 
velocity and similar plots for the E871 conventional drift 
chambers with Ar-\ethane. We observed a drift velocity exceeding 
100\,$\mu$m/nsec, twice that of Ar-\ethane 50:50. This factor 
combined with the improved segmentation ratio reduced the 
individual drift cell occupancy by a factor of eight over the 
conventional wire chambers of E791, the predecessor to E871.

\subsection{Experience with prototypes}

Prototype chambers were built at The University of Texas and Stanford and tested 
both in cosmic-ray telescopes and in the AGS neutral beam. It 
was during these tests that aging effects due to gas 
contamination began to appear.
	
\subsubsection{Sulphur contamination}

After discovering a blackening effect in straws subjected to ethane 
flow during the 1994 engineering run, we analyzed the interior 
surfaces of some of the exposed straws with Auger Electron Spectroscopy 
and a Scanning Tunneling Microscope. We found evidence of sulfating 
of copper due to sulfur contaminants in our ethane supply, and 
concluded that it was necessary to develop a gas filtration system 
for our copperized straw system.

Compounds such as H$_2$S and SO$_2$ are very reactive with copper, 
forming a black copper sulfate which in time replaces the copper, 
destroying the electrical conductivity of the straw and effectively 
removing the cathode. To solve this contamination problem, we 
specified a sulfur content of 0\,ppm for our ethane supply, 
a severe restriction. In addition, we installed high-pressure 
catalyst and sieve filters directly on the ethane gas bottles, 
prior to pressure regulation.

\subsubsection{Effects due to \freon}

Impurities in our freon supply were also found to cause straw 
aging. With a small Mylar-straw test stand, we observed that 
with very low flow rates of freon-based gases the conductive 
copper surface may be removed entirely (`etched') from the 
Mylar surface. 

We observed an etching effect in a controlled experiment using 
a small test chamber consisting of two identical, 60-cm copperized 
Mylar straws. After bringing the straws to an operating voltage 
of 2.3\,KV with \freon-methane (30:70), we halted the gas flow 
and left the bias on one of the straws. This straw began to 
draw current after about 1.5 hours, and at 2 hours was drawing 
5\uA. Meanwhile the copper was disappearing from this straw, 
leaving clear Mylar. A similar test with Ar-ethane (50:50) 
showed no effect. To isolate the problem gas, the test was 
repeated with 100\% \freon; the straw drew 60\uA within 
5 minutes after flow was stopped, eventually leveling off 
at 140\uA. After about 8 hours the copper in this straw was 
visibly thinner.

We have found no reports of any other observations of this effect. 
The most consistent mechanism is this: water vapor entered 
the tubes at a constant rate and with the gas flowing 
at some minimum value, it was flushed out. When the gas flow 
was stopped the water concentration in the straws built up. 
The hydrogen in the water reacted with some \freon 
radicals to make hydrofluoric acid which attacked the copper. 
As a result of this study we increased our gas flow from three 
volume exchanges per day to ten.

\subsection{Gas purification}

\begin{table}
    \caption{Impurities in Semiconductor Grade \freon. Numbers declare the 
    maximum quantity of each impurity in the delivered gas, 
    in units of molar ppm. \label{cf4impurities}}
    \begin{center}
        \begin{tabular}{rc} \hline\hline
                    Impurity & Maximum Quantity \\
                             &  (molar ppm)     \\ \hline
                      CO$_2$ & 10               \\
                         CO  & 5.0              \\
                    H$_{2}$O & 1.0              \\
                       N$_2$ & 200              \\
                  O$_2$ + Ar & 40               \\
           other halocarbons & 10               \\
                      SF$_6$ & 5.0              \\
             total fluorides & 0.1 (ppm weight) \\ \hline
        \end{tabular}
    \end{center}
\end{table}

The gases used were standard commercial products, including 
CP-grade ethane (99.0\% pure) and semiconductor-grade 
\freon (99.7\% pure). Table~\ref{cf4impurities} lists 
impurities found in the \freon as supplied. As has already 
been observed, some of these contaminants can play a role 
in the acceleration of aging effects. Water, oxygen, and 
any halocarbons were the most undesirable impurities. 


\begin{figure*}
    \centerline{\includegraphics[bb=70 490 520 700]{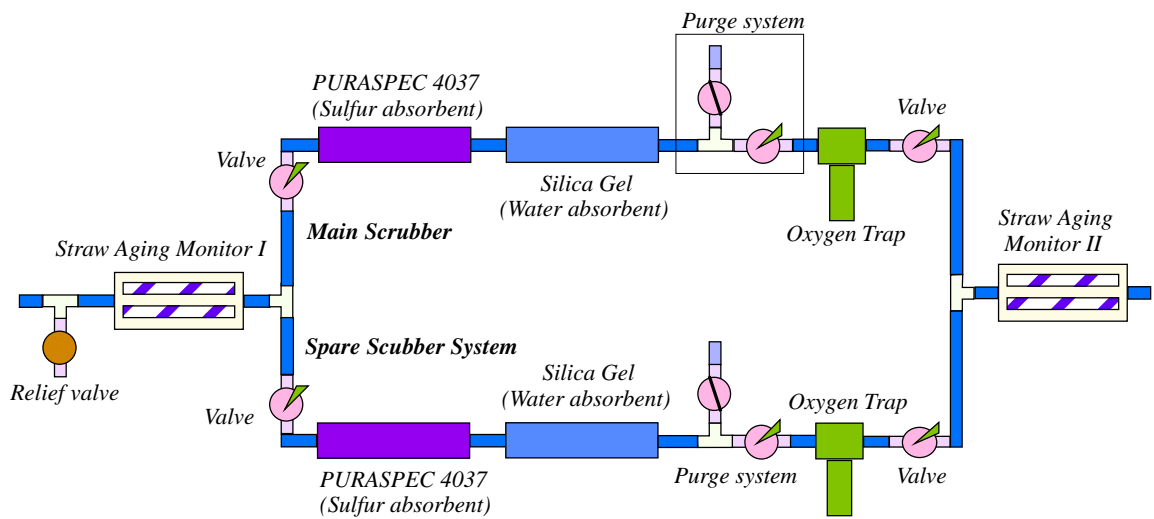}}
    \caption{Low pressure filtration system.} \label{scrubber}
\end{figure*}

A low pressure filtration system was placed after blending 
the gas mixture components. Its goal was to eliminate sulfur, 
water, and oxygen from the gas mixture.

Ethane, derived from natural gas, can contain many parts 
per million (ppm) of H$_2$S, depending on its source. 
While mass-spectrographic analysis indicated that the 
H$_2$S level in our ethane supply was not very high, 
even contamination at a few ppm could destroy our thin 
(1000\,\AA) copper cathode. We therefore developed 
a `scrubbing' system to remove the inevitable sulfur 
contamination. 

We required a simple, safe, inexpensive, self-indicating 
system, which ruled out many of the processes common to 
the petrochemical industry. However the catalyst PURASPEC 4037, 
a material produced by ICI Catalysts of Billingham, 
England~\cite{src:lp-scrub} allows for low-pressure, 
low-temperature elimination of H$_2$S from the gas stream. 
The catalyst is in the form of 0.1-0.2\,in diameter 
spheres composed of 40-45\% copper oxide and 20-30\% zinc 
oxide by weight, with the balance being aluminum oxide. 
These metal oxides react with H$_2$S to produce metal 
sulfides and water vapor. The water vapor must then be 
removed by a second trap, which consisted of silica 
or alumina gel. 

The scrubbing system thus took shape as two 8-in long, 
1.5-in diameter glass cylinders, one filled with PURASPEC 
4037 and a second with silica gel. The cylinders were 
tapered at each end to allow connection to the rest of 
the gas system, while the absorbent materials were 
contained within the cylinders between plugs of glass wool. 
Glass cylinders were used to take advantage of the self 
indicating feature of the absorbents, since PURASPEC 4037 
turns from gray/green to black when loaded with sulfur, 
while alumina gel turns from blue to pink when saturated 
with water. We included a redundant scrubbing line to 
prevent disruption during absorber replacement. Heavy-walled 
glass was used throughout. To prevent overpressure, we 
installed a check valve with a crack pressure of 20\,psi
before the scrubber system. An oxygen trap~\cite{src:ox-trap} 
completed the scrubber system.

Immediately upstream and downstream of the scrubber system, 
we installed reaction monitors, consisting of clear 
plexiglas tubes containing copper straws. They were used 
as indicators of chemical reactions uncorrelated to the 
radiation environment of the chambers. Figure~\ref{scrubber} 
depicts the complete low-pressure scrubber system. 


\begin{figure}
    \centerline{
    \includegraphics[bb=0 0 384 449,width=.4\textwidth]{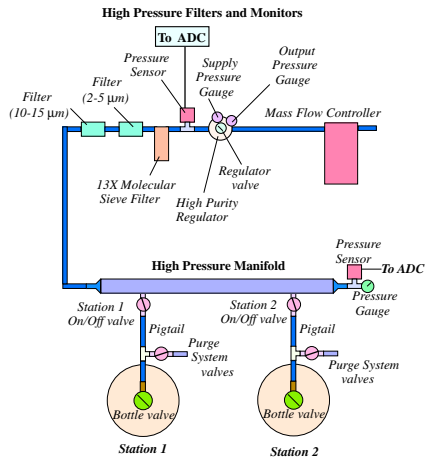}}
    \caption{High pressure manifold and filtration system.} 
    \label{hp_filters}
\end{figure}

In addition to the low-pressure scrubber system, we installed 
a high pressure filtration system directly on the ethane 
bottles to remove oil, water, and possible large debris 
upstream of the high-pressure regulators. The system was 
composed of a pair of mechanical filters effective at 
eliminating contaminants greater than 10-15 and 2-5 microns 
respectively, available from many sources in varying 
form~\cite{src:hp-filt}. The final stage of the high pressure 
filtration system was a type 13X molecular 
sieve~\cite{src:hp-sieve}, a crystaline metal 
aluminosilicate. This type of sieve material is 
composed of Na$_2$O, Al$_2$O$_3$, and SiO$_2$. 
It is capable of removing H$_2$0, CO$_2$, H$_2$S, 
oil, and metacaptans (compounds with an alcohol-like 
structure, with the OH radical replaced by an SH radical). 
The nominal pore diameter is 10\,\AA, and expected 
levels of H$_2$S after this high pressure system were 
below 0.1\,ppm.

\subsection{Flow control}

Drift gas mixtures and flow rates were controlled by a Matheson 
Multichannel Dyna-Blender~\cite{src:dynablender}. Each channel 
was a mass flow controller that delivered the appropriate flow 
near atmospheric pressure, regardless of the input pressure. 
The blender maintained gas mixture ratios to 2\% precision at 
operational flow rates, and adjusted as necessary for fluctuations 
in input pressure. It simultaneously provided online flow readings, 
so it also served as an indicator of gas-supply shortages. 

Total helium flow to the volumes surrounding the straws was 
controlled by a dependent channel of the blender. This channel 
was maintained at some fraction of the flow rate of the total 
internal straw flow, thus limiting the differential pressure 
experienced by the straws. 

The blender regulated the total flow of the gas mixture and 
the helium. Distribution of the gas to the 8 detectors 
(14 modules) was accomplished by a parallel set of rotameters. 
Vent bubblers were used to prevent contamination of the gas 
volume from back flow and maintained a minimum pressure 
inside the chamber. Differential pressure sensors~\cite{src:p-sens} 
monitored the difference in pressure between the drift gas 
and surrounding helium volume at each chamber. 

\subsection{Helium volumes}


To minimize multiple scattering, polyurethane bags filled 
with helium surrounded the whole detector, filling the spectrometer 
fiducial volume. Helium also filled the volumes made up 
by chamber frames, outside the straws. The gas system was 
tuned to maintain a positive outward pressure gradient on 
the straw tube walls. Higher pressure was needed on the 
inner surface since the straw tubes were very sensitive 
to outside forces that were not compensated from the inside. 
Lab tests had shown that straws could withstand at least 
20\,psig of inner pressure without any damage. On the other 
hand, the smallest pressure on the outside surface deformed them. 
If this deformation was more than 1-2\,cm long, the cathode did 
not return to its original shape; the resulting defect could 
thus distort the straw electric field. Helium was flowed through 
the volumes rather than statically filling them, to prevent 
the potential accumulation of flammable gasses.


As a final precaution, a set of differential pressure sensors were installed.
They continuously measured the differential pressure between the inside and the
outside of the straws. A pressure threshold of 2.54~cm of water  was used to generate an
alarm. Readings of the differential pressures were  displayed online in the
control room.

The helium supplied to the helium bags surrounding the chambers was independent
from the helium supply inside. A system of photohelic gauges monitored the
differential pressure of the  inside and the outside of the bags. Their mission
was to automatically switch on and off actuator valves so that the bags were
always full. If the pressure had exceeded a prespecified pressure level, then
the  bag would start pushing the thin Mylar windows enclosing the chambers'
helium volume. The bags would have effectively applied an increased pressure on
the outside of the straw tubes. Relief valves (bubblers) were placed at the
output of the helium bags to protect the chambers in case of failure of the
photohelic system.

\begin{figure*}
\centerline{ \includegraphics[bb=0 190 437 500]{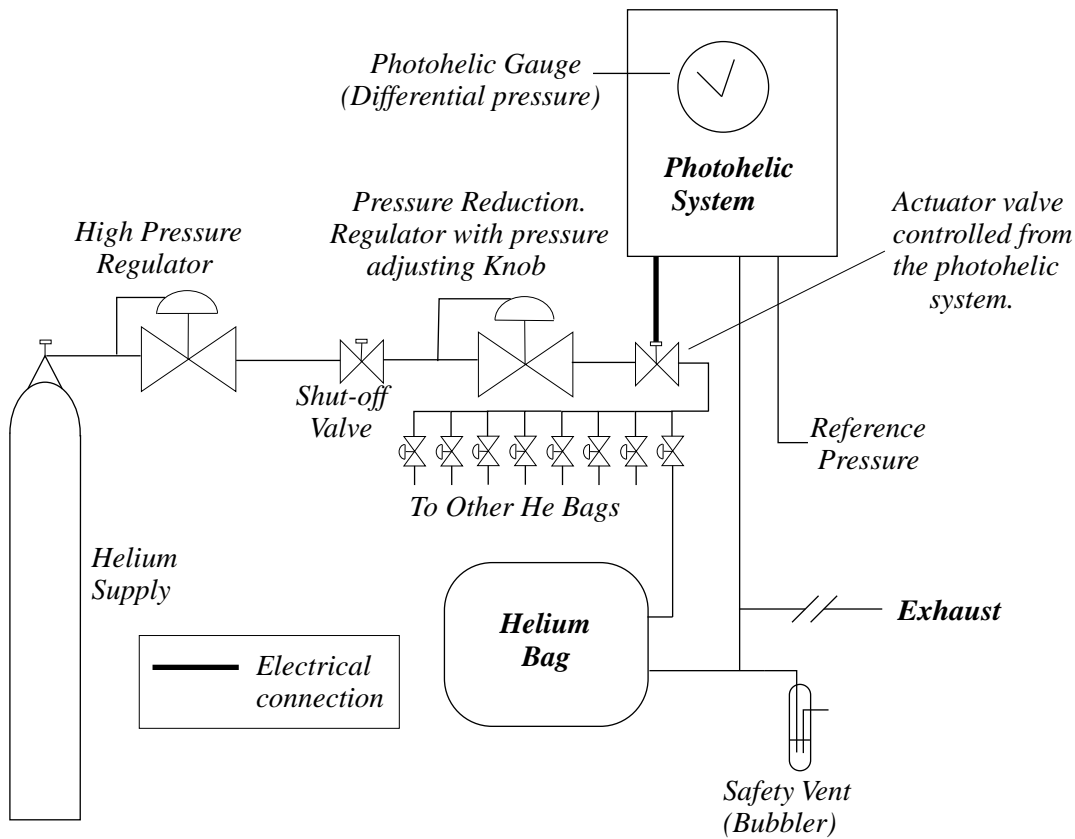} }
\caption{Helium system} \label{photohelic}
\end{figure*}

\subsection{Operating conditions}

\begin{table}
  \caption{He volumes and flow rates, by module. Helium gas volume 
  (all modules) : 1871.9\,L, Nominal Helium flow: 2050-2300\,ccm 
  (1.5-1.8 volume exchanges/day).} \label{he_flow}
  \begin{center}
    \begin{tabular}{rcccc}
      \hline
      \hline
            & \multicolumn{4}{c}{Station} \\
            & 1 & 2 & 3 & 4 \\
      \hline 
      height (in)       & 24.077 & 34.472 & 54.904 & 57.688    \\
      width (in)        & 8.600  & 8.600  & 8.600  & 8.600     \\
      length (in)       & 37.075 & 37.775 & 40.282 & 38.740     \\
      volume (L)        & 125.8 & 183.5 & 311.7 & 314.95 \\
      \hline
    \end{tabular}
  \end{center}
\end{table}
\begin{table}
  \caption{Drift gas volumes and flow rates, by module. 
  	Total Straw Tubes Gas volume: 122.7\,L, Nominal Gas 
	Mixture flow: 800-960\,ccm (9.4-11.3 volume exchanges/day). 
	Single straw cross section: 0.19662\,cm$^2$ }
	   \label{gasvol_flow}
  \begin{center}
    \begin{tabular}{rcccc}\hline\hline
      & \multicolumn{4}{c}{Station} \\
      $x$ view & 1 & 2 & 3 & 4 \\
      \hline
      length (cm) 	   & 80.36  & 89.40 & 113.97    & 119.67 \\
      No. straws  	   & 513    & 516   & 550       & 531  \\
      volume (L)           & 8.11   &  9.07 & 12.32     & 12.07\\
      \hline
      $y$ view & 1 & 2 & 3 & 4 \\ \hline
      length (cm) 	   & 88.88  & 89.03 & na    & 91.85 \\
      No. straws   	   & 309    & 344   & na    &  463 \\
       volume (L)          & 5.40   & 6.02  & na    & 8.36\\
       \hline
    \end{tabular}
  \end{center}
\end{table}

Keeping down the cost of our \freon supply meant restricting the flow 
of our gas mixture to the lowest safe value. Due to its small molecular 
size, helium could invade the straw volume through the straw walls, 
affecting mixture purity. In addition, electron impact dissociation in 
the gas produces F$^-$ and CF$_3^-$ negative ions together with F$^*$, 
CF$_2^*$ and CF$_3^*$ radicals. Too little flow would have allowed these 
radicals enough time to become involved in polymerization and further 
chemical reactions, including etching. The copper layer is extremely 
volatile to fluorine or hydrogen fluoride (HF) under the proper conditions 
as a non-dry environment. 

During the first running period of the experiment, the flow was set 
to 3-5 volume exchanges per day. Lab tests had shown no problems 
with this flow rate. However, these had been performed outside of the intense 
radiation environment of the E871 spectrometer, and not for such long periods. 
As has been discussed, we eventually discovered signs of etching in 
the most heavily radiated straws. For the 1995 run period the flow was 
increased to 8-10 volume exchanges per day, freezing visible etching 
effects. In the new set of chamber modules used in the 5-month 1996 
run period, we observed no etching effects with this higher flow rate.

Gas volumes and flow exchanges are listed in Tables~\ref{he_flow} 
and~\ref{gasvol_flow}.


\subsection{Running experience}

The gas system operated successfully, providing the necessary 
purity of gases while maintaining a constant flow and mixture 
ratio. No sulfur contamination was detected and no aging effects 
were observed in the performance of the detector. No further 
evidence of aging was observed after the flow rate increase and 
installation of the scrubbing systems.

During the 12 months of operation the PURASPEC 4037 material had 
slightly changed color, indicative of sulfur trapping, while the 
silica gel was twice overloaded with water. The NANOCHEM oxygen 
cartridge was very sensitive, easily overloaded with oxygen in 
open air contact, and without self-indication. We replaced it on 
a regular schedule (after every second cylinder of \freon used), 
but the system was always vulnerable to contamination from 
insufficient purging upstream or from any leakage downstream. 


The differential pressure sensors proved to be extremely useful. With gas
lines of tens of meters lengths and numerous connections, valves and distribution
manifolds, it was easy to limit by accident either the input or the output flow
of the gas mixture. At least twice they provided critical alarms of higher
pressure on the outside helium volume. The pressure was sensed through long
lines since the actual sensors were located far away from the detectors
(outside the high radiation area). This caused a delay in measuring the
actual pressure gradient.

The use of manifolds with multiple stations, double filtering systems and
purging capabilities, allowed the continuous and uninterrupted flow of gas for
long periods (6 months). The pressure and flow sensors provided
warnings for each irregularity in the normal operation and reminders for the
necessary maintenance of the system. The precision of each gas component flow was
$\pm$10\,ccm, resulting in a 5\% uncertainty in the mixture ratio at the
operational flow. The distribution of flow to individual modules was done with
flowmeters with flow adjusting valves. These provided only
a rough measurement of the portion of the total flow directed to each module.
They were unreliable and two of them had to be replaced because of
malfunction. Also, the Matheson high pressure sensors had a high rate of
failure; two stopped working after months of normal operation.



\section{Performance}

\subsection{Chamber calibration}

\subsubsection{Alignment}

The straw chambers were installed in the E871 spectrometer on tables of 
aluminum jig plate, which had been surveyed relative to the beam axis 
and target. The chambers were pinned and bolted to the tables before 
they themselves were surveyed. Much more accurate relative positions 
information could be obtained using track data recorded with the spectrometer 
magnets off. Corrections to the optical survey results were derived 
from this straight-through data. 

Relative transverse corrections to chamber positions within a view ($x$ or $y$,
left or right) can be readily extracted from straight-through tracks, but
longitudinal (beam parallel) and global corrections require more work. The relative positions
of the left and right arms of the spectrometer were extracted from reconstructed
vertex distributions, while longitudinal corrections within
spectrometer arms were derived from magnet-on data, where allowances for multiple
scattering in the track fitting algorithm had been disabled in software.

Alignment was performed iteratively. The tracking $\chi^2$, measured with an
independent fitting algorithm, reduced systematic errors from the combined effect
of the magnetic field map and the spectrometer survey to below 1\%.

\subsubsection{Timing and drift velocity}

For every change in straw and drift chamber operating conditions, we 
generated new calibrations for TDC time offsets, single-wire efficiencies, 
and effective time-to-distance relations. More than sixty calibration updates 
were needed for the two years' run periods.

Normal minimum-bias data and random-trigger data were used to perform 
the calibrations, which were done in an iterative fashion. Noise rates 
determined hot or dead wires, while time sum residuals between neighboring wire pairs 
were minimized to determine the TDC time offsets.

\subsection{Straw system performance}
	
\begin{figure*}
  \centering
  \includegraphics[width=.8\textwidth]{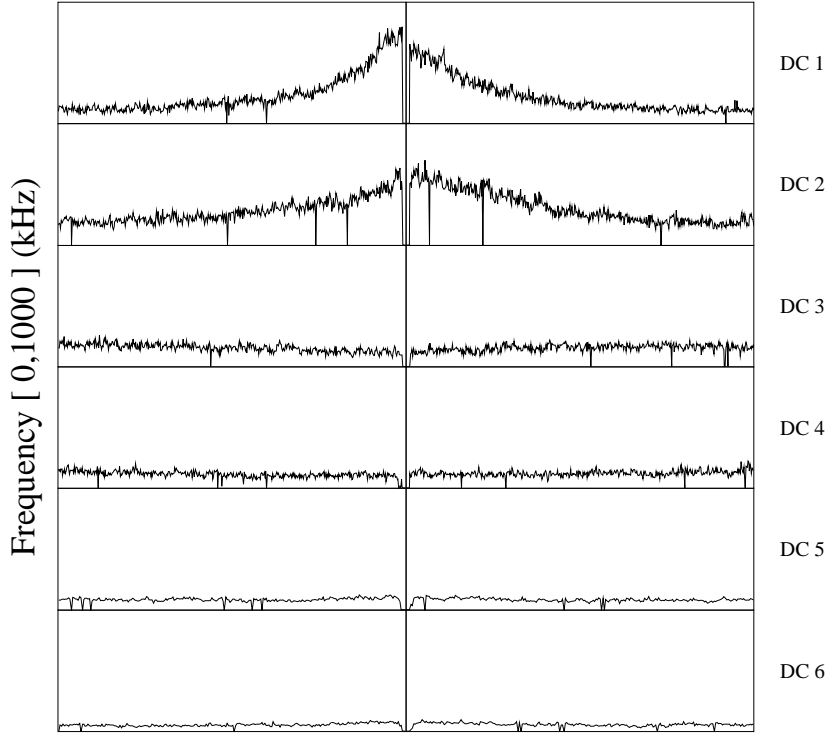}
  \caption[Tracking chamber hit rates]{Tracking chamber hit rates.  Hit rate
  vs. wire number for a random trigger at 15\,Tp.}\label{f:dc-rates}
\end{figure*}

\begin{figure}
  \centerline{\includegraphics[bb= 0 20 283 240]{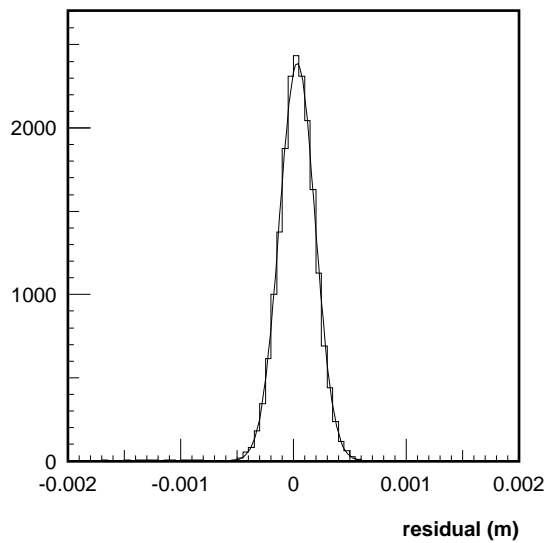}}
  \caption{Typical single-wire residual (in meters) for a straw chamber. 
  The width is $158.5\pm0.9\um$.}\label{f:sdc-res}
\end{figure}
  
\begin{figure}
  \centerline{\includegraphics[bb= 0 20 283 240]{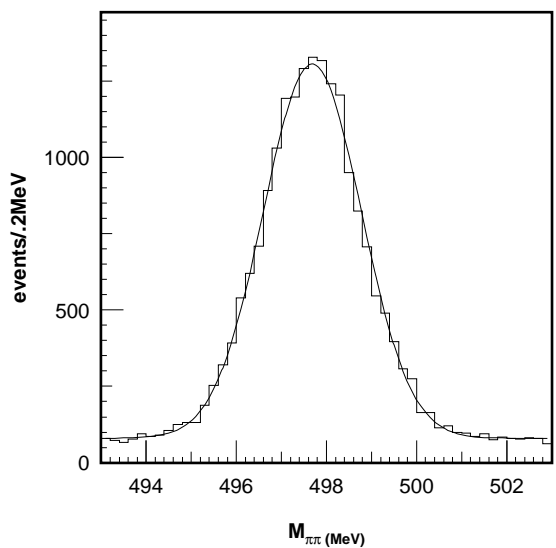}}
  \caption{Reconstructed kaon mass for \KPP data.
  Two-body decays were selected by rejecting events with more
  than $10\MeV/c$ momentum component perpendicular to the beam
  direction, $p_t$. 
  Our mass resolution for \KPP was 1.11\MeV.}
  \label{f:pipi-mass}
\end{figure}

As shown in Figure~\ref{f:dc-rates}, noise rates for individual wires 
in the straw chambers approached 750\,kHz at 15$\times 10^{12}$ protons
(Tp) on target for wires near 
the beam and in front of the beam stop. The continuous cathodes and 
stability of the straws allowed smaller cell size for decreased 
occupancy, while minimizing crosstalk. 
A typical residual distribution is shown in Figure~\ref{f:sdc-res}. 
The mass resolution of the spectrometer as a whole is shown in 
Figure~\ref{f:pipi-mass}, which shows a \KPP mass peak with 
a $\pi^+\pi^-$ mass resolution of 1.11\MeV.

\begin{figure*}
  \centerline{\includegraphics{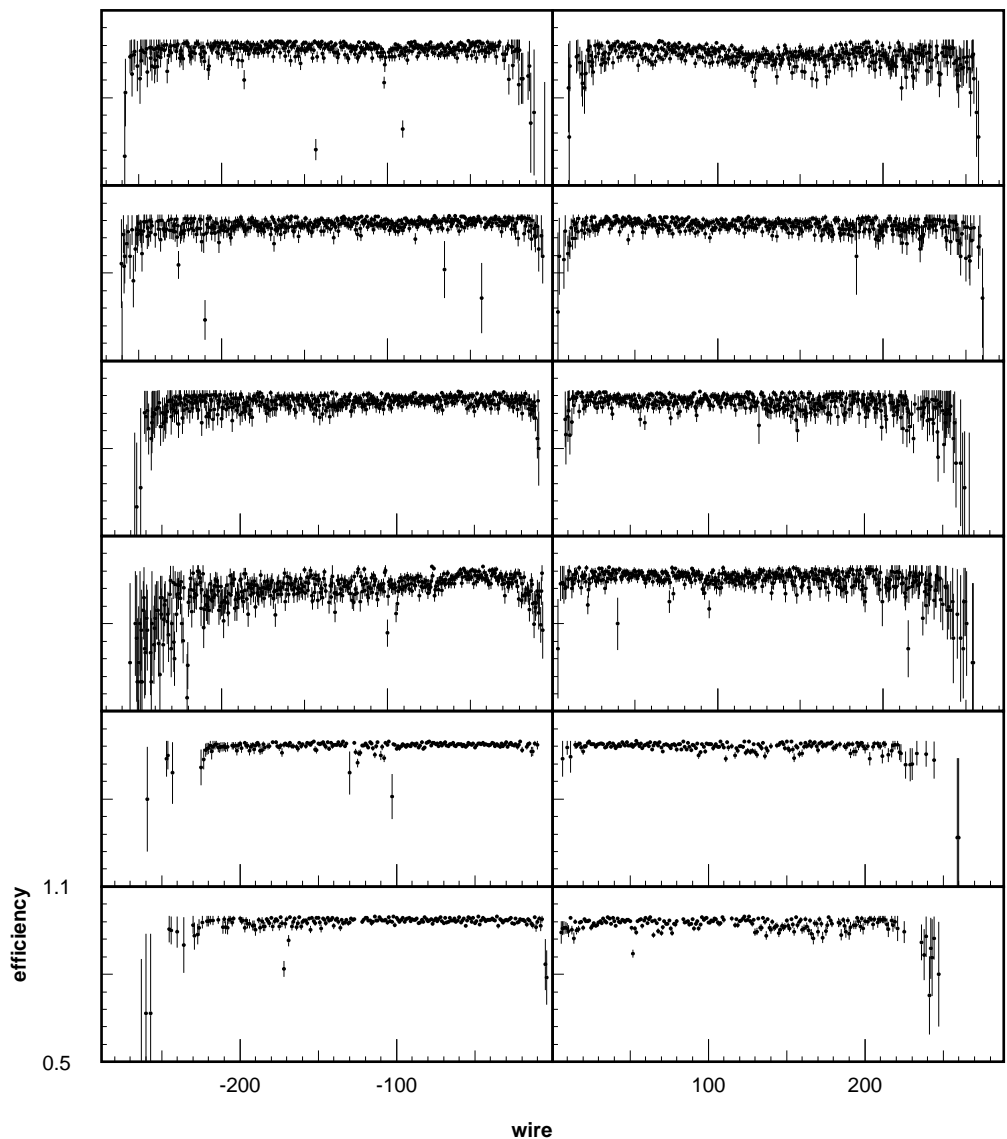}}
  \caption{Single-channel efficiency for the E871 straw and 
  conventional drift chambers, by channel and chamber (top to bottom: 
  SDC1-4 and DC5-6. Straw chambers have roughly twice as many 
  wires as drift chambers).}
  \label{f:eff-array}
\end{figure*}

\begin{figure}
  \centerline{\includegraphics[bb= 0 20 283 240]{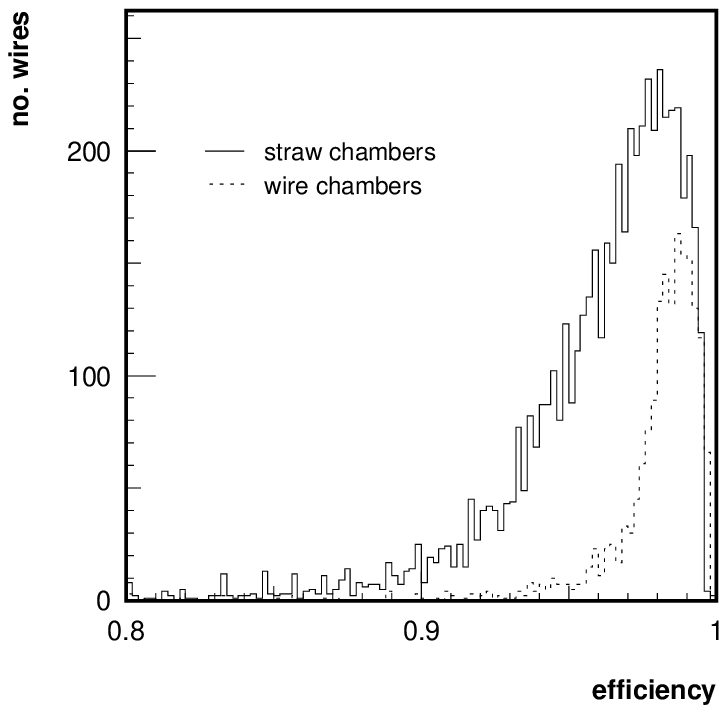}}
  \caption{Single-channel efficiency for the E871 straw and 
  conventional drift chambers.}\label{f:eff-wire}
\end{figure}

Straw efficiency was flat across most of the cell area, 
dropping off near the straw wall, where particles traversed increasingly 
smaller gas path lengths. The flatness of the efficiency as a function 
of wire number (and thus position) for the most upstream chamber 
(for example), shown 
in Figure~\ref{f:eff-array}, clearly displays the stability of the E871 
straw chambers regardless of occupancy, since, as evident from 
Figure~\ref{f:dc-rates}, the upstream chamber intensity varies across the chamber 
face. Figure~\ref{f:eff-wire} displays the single-channel efficiencies 
for all straw and conventional drift chambers. The mean efficiencies 
for straws and drift chambers were 96\% and 98\%, respectively.

\section{Summary}

We successfully developed, built, and operated a set of straw tracking
chambers with over 
6500 channels for the BNL E871 rare decay search. The system proved to 
be robust, with no rate-induced tripping or inefficiency at rates over 
700\,KHz. The system utilized
 a close-pack cell 
configuration of Mylar or Kapton 0.5-cm diameter, copper-lined straws 
with a 50:50 \freon-\ethane drift gas at atmospheric pressure. We 
developed custom front-end amplifiers and electronics for the chambers, 
and operated them for two years' running periods at 
Brookhaven National Laboratory.

 

\begin{thebibliography}{99}

\bibitem{e871:98ue}
    D. Ambrose \etal (BNL E871 Collaboration), \emph{Phys. Rev. Let.} \textbf{81}
    (1998) 5734.
 
\bibitem{e871:98ee}
    D. Ambrose \etal (BNL E871
    Collaboration), \emph{Phys. Rev. Let.} \textbf{81} (1998) 4309.
 
\bibitem{e871:98uu}
    D. Ambrose \etal (BNL E871 Collaboration), \emph{Phys. Rev. Let.} 
    (in preparation).
 
\bibitem{worm:95}
    S. Worm, Ph. D. Thesis, University of Texas, Austin, June, 1995.
        
\bibitem{belz:98}
    J. Belz \etal, 
    \emph{Nucl. Inst. Met.} (submitted, e-print hep-ex/9808037).

\bibitem{VertexStraws}
    M. Frautschi \etal (AMY Collaboration), \emph{Nucl. Inst. Met.} A307 (1991) 52-62.

\bibitem{BaloonStraws}
    A. Tomasch \etal, in Adelaide 1990, Proceedings, Cosmic ray, vol. 4* 414-417.

\bibitem{AtlasStraws}
    Bing Zhou \etal, IEEE Trans.\ Nucl.\ Sci.\  {\bf 37}, 1564 (1990).
    V.~Bondarenko {\it et al.},
	Nucl.\ Instrum.\ Meth.\ A {\bf 327}, 386 (1993).
	%
	V.~Commishau {\it et al.},
	CERN-DRDC-93-46.
	%
	T.~Akesson {\it et al.},
	Nucl.\ Instrum.\ Meth.\ A {\bf 361}, 440 (1995).
	%
	T.~Akesson {\it et al.}  [ATLAS-TRT Collaboration],
	Nucl.\ Instrum.\ Meth.\ A {\bf 449}, 446 (2000).

\bibitem{tsai:74}
    Y. S. (Paul) Tsai, \emph{Rev. Mod. Phy.} 46 (1974) 815.

\bibitem{mylar}
    Mylar is DuPont Corporation's brand name of a polyester film.

\bibitem{kapton}
    Kapton is DuPont Corporation's brand name of a polyimide film.
    
\bibitem{Majewski_polycarbonate}
    T.S. Shin {\it et al.}, \emph{Nucl. Inst. Met.} \textbf{A332} (1993) 469.
	
\bibitem{straw-manu}
    The Mylar straws for 1995 were wound by Stone Industrial, 
    9207 51st Ave., College Park, MD 20740-1910, (301)474-3100.

\bibitem{straw-manu-2}
    The Kapton straws straws for 1996 were wound by Lamina Dielectrics Ltd.,
    Myrtle Lane, Billingshurst, West Sussex, RH14 9SG, UK,
    +44 1403 783131.
    
\bibitem{wire-manu}
    20\um, gold-plated tungsten wire supplied by SAES Getters, 
    1122 E. Cheyenne Mtn. Blvd., Colorado Springs, CO 80906, (719)576-3200.

\bibitem{vapor-dep}
    The vapor-deposited Mylar and Kapton was supplied by 
    Sheldahl Corp. P.O. Box 170,
    Northfield, MN 55057, (507)663-8000.

\bibitem{ultem-manu}
    Ultem 1000 is a registered trademark name for polyetherimide resin by
    General Electric Company. Ours was supplied by Polymerland, 
    500 Park Blvd., Suite 245, Itasca, IL 60143, (800)752-7842.
  
\bibitem{Tyvek}
    Tyvek is a product of DuPont Corporation.

\bibitem{McDonald_straw_cutting}
    K. McDonald {\it et al.}, Princeton University, private communication.
    
\bibitem{src:feedthru}
    Feedthrough molding by Advance Tool and Die, 226 Highland Ave., 
    Westmont, NJ 08108, (609)854-6329.
    
\bibitem{src:sleeve}
    Brass sleeves and pins machined by J\&J Swiss Precision, Inc., 
    160 West Industry Court, Deer Park, NY 11729, (516)243-5584.
    
\bibitem{src:frame}
    Frames machined by Brigg's Machining Co., 23190 Del Lago, 
    Laguna Hills, CA 92653, (714)770-1160.
    
\bibitem{epo-tek}
    EPO-TEK 301-2, EPO-TEK 110, and EPO-TEK 410E were supplied by 
    Epoxy Technology Inc., 14 Fortune Dr., 
    Billerica, MA 01821, (508)667-3805.
    
\bibitem{tra-con}
    Tra-Bond 2143D supplied by Tra-Con, Inc., 55 North St., 
    Medford, MA 02155.

\bibitem{src:encap}
    Dow 3110-RTV by Dow Corning, Midland, MI 48640, (800)248-2481.

\bibitem{src:disp}
    EFD Dispensing Components, 977 Waterman Ave., East Providence, 
    RI 02914, (800)828-3331.
    
\bibitem{fnal831}
    M. Hamela and P. Yager, private communication.
    
\bibitem{WireTensionMonitor}
     K. Lang, J. Ting, and V. Vassilakopoulos, 
     \emph{Nucl. Inst. Met.} \textbf{A420} (1999) 392.
    
\bibitem{graessle:95}
    S. Graessle \etal,
    \emph{Nucl. Inst. Met.} \textbf{A137} (1995) 138.
    
\bibitem{heinson:92}
    A. P. Heinson and D. Rowe, 
    \emph{Nucl. Inst. Met.} \textbf{A321} (1992) 165.
    

\bibitem{src:lp-scrub}
The U.S. representative for ICI Catalysts was ICI Katalco, 
Two Trans Am Plaza Dr., Suite 230, Oakbrook Terrace, 
IL 60181, (708) 268-6300.

\bibitem{src:ox-trap}
Oxygen trap with replaceable cartidge, Scott Specialty Gases, Part no.
53-42CT. Each catridge will remove 99\% of the oxygen present in a gas stream
with a 15ppm O$_2$ level. Its capacity  is 127 ml of O$_2$ at 99\% efficiency.

\bibitem{r:fast-tdc}
    %
    R.~D.~Cousins, C.~Friedman and P.~L.~Melese,
    IEEE Trans.\ Nucl.\ Sci.\  {\bf 36}, 646 (1989).
    %
    R.~D.~Cousins {\it et al.},
    Nucl.\ Instrum.\ Meth.\ A {\bf 277}, 517 (1989).


\bibitem{Vavra:1986sy}
J.~Vavra,
Nucl. Instrum. Meth. {\bf A252}, 547 (1986).

\bibitem{Vavra:1993cy}
J.~Vavra, P.A.~Coyle, J.A.~Kadyk and J.~Wise,
Nucl. Instrum. Meth. {\bf A324}, 113 (1993).


\bibitem{src:hp-filt}
 High pressure gas filter, Scott Specialty Gases Part no. 
 53-45F-111 (53-45F-112) or Hoke Part No. 6321F4B (6323F4B). 
 Max pressure 3000\,psi, removes particles larger than 2-5 
 (10-15) microns.  
 
\bibitem{src:hp-sieve}
High Pressure filter, Scott Specialty Gases, Part no. 53-43H, Max.
Pressure 3000\,psig. Brass with Viton O-ring. Active
purifier element: Molecular sieve 13X (Part no. 53-43E); removes oil,
water and particulates.

\bibitem{src:dynablender}
 Our gas flow controller console was the Matheson Dynamic Gas 
 Blending System Series 8284 Dyna-Blender, a system equipped 
 with four independent transducers/controllers, Matheson Part 
 No. 8272-0413,-0424. They were calibrated for \freon, \ethane, 
 isobutane, and Helium with input pressure of 20\,psig and 
 atmospheric output pressure. 
 
\bibitem{src:p-sens} Pressure Transducers, Matheson part No. 8816
(Range 0-3000 psia).

\bibitem{src:hp-reg}
  Dual Stage High Purity Brass Regulator, Matheson Model 3120 Series, with
  Teflon seats. It was used for \freon, Ar gases. Matheson Model 3100 Series
  was used for hydrocarbons.  

\end{thebibliography}
\end{document}